\documentclass[aps,prb,twocolumn]{revtex4-2}
\usepackage{color}
\usepackage{graphicx}
\usepackage{cancel}
\usepackage{amsfonts}
\usepackage{amssymb}
\usepackage{amsthm}
\usepackage{ulem}
\usepackage{url}
\usepackage{hyperref}
\usepackage{multirow}  % multirow tabular
\usepackage{bm}        % bold math
\usepackage{enumerate}
\usepackage{mathtools}

\usepackage[dvipsnames]{xcolor}
\hypersetup{
	colorlinks,
	linkcolor={green!50!black},
	urlcolor={blue!80!black},
	citecolor   = {red!50!black} %Colour of citations
}
\begin{document}

\title{Bilinear magnetoresistance in 2DEG with isotropic cubic Rashba spin-orbit interaction}

\author{A. Krzyżewska, A. Dyrdał}
%\email[]{ppigon@agh.edu.pl}
%\thanks{}
\affiliation{
Department of Mesoscopic Physics, ISQI, 
Faculty of Physics, Adam Mickiewicz University in Pozna\'n, ul. Uniwersytetu Pozna\'nskiego 2, 61-614
Pozna\'n, Poland
}

\begin{abstract}
Bilinear magnetoresistance has been studied theoretically in 2D systems with isotropic cubic form of Rashba spin-orbit interaction. We have derived the effective spin-orbital field due to current-induced spin polarization and discussed its contribution to the unidirectional system response. The analysed model can be applied to the semiconductor quantum wells as well as 2DEG at the surfaces and interfaces of perovskite oxides.
\end{abstract}

\maketitle

%\tableofcontents

\section{Introduction}
\label{sec:introduction}
%The magnetotransport
%is a well-known since over a hundred years in condensed matter physics ~\cite{Campbell1923Jan,ziman_1972,Pippard1989Sep}. 
The magnetotransport
is a phenomenon that is know in condensed matter physics for more than hundred years~\cite{Campbell1923Jan,ziman_1972,Pippard1989Sep}. 
The associated magnetoresistance is currently also a hallmark of spin electronics, that was developed after the   discovery of giant magnetoresistance in metallic multilayers at the end of 80's ~\cite{Grunber_GMR,Fert_GMR,Camley}.  Nowadays spintronics takes advantage of spin-to-charge interconversion effects~\cite{Sinova_RevModPhys2015,Ando_JPSJ_rev,Soumyanarayanan,Wu_spin-charge} that lead to new  magnetoresistance phenomena such as spin magnetoresistance and various unidirectional magnetotransport effects~\cite{Chen2013,Nakayama2013,Kim_2016,Avci2015,Lv2018,Tokura2018}. 

The magnetoresistance (MR) is typically a quadratic function of the magnetic field  (or magnetization) amplitude. The linear scaling  with magnetic field/magnetization is rather unique and dictated by specific symmetry requirements. Recently, the linear dependence on the magnetic field has been reported in single crystals of antiferromagnetic metals (i.e., TmB4), where  MR can be tuned from quadratic to linear one depending on the orientation of magnetic field~\cite{Mitra2019Jan,Feng2019Jun}.
It has been shown in recent years, that the spin-orbit interaction can lead to spin-currents and non-equilibrium spin polarization, that can coupled to the external magnetic field or equilibrium magnetization of the system leading to unidirectional magnetotransport~\cite{He2018May,Zhang2018Aug,Zhang2018Sep,Dyrdal2020Jan,Vaz2020Jul}. The unidirectional spin magnetoresistance in general is a consequence of non-equilibrium spin polarization at the interface of the hybrid structures consisting of a ferromagnetic thin film and a layer of heavy metal or topological insulator (TI). Interestingly, it has been shown that the unidirectional magnetoresistance also appears in nonmagnetic materials. This effect is called bilinear magnetoresistance, BMR, as it scales linearly with the charge current density and external magnetic field. 
The BMR effect can be a consequence of the strong spin-orbit interaction in systems with highly anisotropic Fermi contours. This is for example a case of TIs with strong  hexagonal warping~\cite{Zhang2018Aug,He2018May}. In this case the nonlinear spin currents appear in the system, which in the presence of external magnetic field are partially converted to the nonlinear charge current. 
Another mechanism was proposed to explain the bilinear magnetoresistance in systems with isotropic Fermi contours. This mechanism is related to the non-equilibrium spin-polarization (also known as inverse spin-galvanic effect or Edelstein effect) that appears in the system under external electric field and leads to the effective spin-orbital field that couples to the electron spin.~\cite{Dyrdal2020Jan}. 
The Berry curvature dipole can also give contribution to the unidirectional magnetotransport. 

In this work we analyze the BMR in the 2D system with isotropic cubic form of Rashba spin-orbit coupling (SOC). Such a form of Rashba SOC can be found in 2DEG emerging in the semiconductor quantum wells as well as 2DEG at the surfaces and interfaces of perovskite oxides~\cite{Moriya2014Aug,Liang2015Aug,Nakamura2012May,Zhong2013Apr,Khalsa2013Jul,Kim2013Jun,Shanavas2016Jan,Seibold2017Dec,Vaz2019Nov,Dyrdal2020Jan}. The detailed description of the model of 2DEG electron gas in the presence of isotropic cubic Rashba SOC  can be found, e.g., in
~\cite{Liu2008Mar,Karwacki2018Jun}.

The paper is organised as follows. In sec.~\ref{sec:model_method} we will present the effective Hamiltonian describing 2DEG with isotropic cubic Rashba coupling and formalism that will be used to describe transport characteristics. Next, in the sec.~\ref{sec:Spin-to-charge_B0}, we will derive analytical expressions for charge current density and non-equilibrium spin polarization in the absence of external magnetic field.
The main results will be presented in Sec.~\ref{sec:magnetotransport}, where we will present derived analytical expressions for  bilinear magnetoresistance. Next, we will present numerical results. The general conclusions and summary will be provided in
Sec.~\ref{sec:summary}.  

%======================================
\section{Model and method}
\label{sec:model_method}
%======================================

%-------------------------------------
\subsection{Effective Hamiltonian}
%-------------------------------------

We consider the effective Hamiltonian, obtained upon two canonical transformations from ${8\times8}$ Luttinger Hamiltonian for p-doped semiconductor quantum wells with structural inversion assymmetry~\cite{LuttingerKohn1955,Luttinger1956}, and takes form ~\cite{Liu2008Mar}:
\begin{equation}\label{eq:hamiltonian}
	\hat{H} = \frac{\hbar^2k^2}{2m^{\ast}} \sigma_0 + i \alpha (k^3_-\sigma_+ - k^3_+\sigma_-) + \mathbf{B}\cdot \hat{\mathbf{s}} + \mathbf{B}_{\mathrm{so}}\cdot \hat{\mathbf{s}}, 
\end{equation}
where $k = \sqrt{k_x^2 + k_y^2}$ is the wavevector amplitude, $m^{\ast}$ is the effective mass. The second therm describes the effect of cubic Rashba SOI~\cite{Liu2008Mar} and is expressed by Rashba coupling parameter, $\alpha$, and  $k_{\pm} = (k_{x} \pm i k_{y})$. In addition, $\sigma_{\pm} = \frac{1}{2}(\sigma_{x} \pm i \sigma_{y})$, where $\sigma_{n}$ ($n=x,y,z$) denotes $n$-th Pauli matrix. Note that, $m^{ast}$ and $\alpha$ are material parameters and are defined in   ~\ref{App:AppA}.
%}
The effect of the external in-plane magnetic field, $\mathbf{B}$, is taken into account by the Zeeman term, where $\mathbf{B} = \{B_x, B_y, 0\}$ is in the energy units, that is $\mathbf{B} = g\mu_B \mathbf{b}$ ($g$ - g-factor, $\mu_B$ - Bohr magneton, $\mathbf{b}$ - magnetic field in Tesla), and spin operators are $2\times2$ matrices that can be written by the identity matrix and Pauli matrices as:
\begin{equation}
\label{eq:spinOp}
\hat{s}_{\alpha} = \sum_{\beta = 0,x,y,z}s^{\alpha}_{ \beta} \sigma_\beta ,
\end{equation}
where the coefficients $s^{\alpha}_{ \beta}$ are provided 
in~\ref{App:AppA}.
\\
Finally, the last term in the Hamiltonian describes the coupling of electron spins to the effective spin-orbital field, $\mathbf{B}_{\mathrm{so}}$, that can be expressed by the non-equilibrium spin polarization emerging due to the inverse spin-galvanic effect and proportional to the charge current density (and thus to the external electric field), i.e, $\mathbf{B}_{\mathrm{so}} \sim \mathbf{S} \sim \mathbf{j}$. As the effective cubic Rashba model considered here is defined by much more complex spin operators, the coupling between spin polarization and electron spin is defined by the two coupling constants: $J_{0}$ that couples non-equilibrium spin polarization to the part of the spin operator proportional to the identity matrix $\sigma_{0}$, and coupling constant $J_{1}$ that couples spin polarization to the part of spin operator proportional to Pauli matrices $\sigma_{x,y}$, 
\begin{equation}
\mathbf{B}_{\mathrm{so}}\cdot \hat{\mathbf{s}} = 
%\check{J} \mathbf{S} \cdot \hat{\mathbf{s}} =
 J_{0} S_{\alpha} s^{\alpha}_{0} \sigma_{0} + J_{1} S_{\alpha} (s^{\alpha}_{x} \sigma_{x} + s^{\alpha}_{y} \sigma_{y})
 \end{equation}
where $S_{\alpha}$ is $\alpha$-th component of the non-equilibrium spin polarization. The coupling constants $J_{0,1}$ have been derived and presented in~\ref{App:Jparameters}.
Here we stress that  non-equilibrium spin polarization induced by the charge current in 2DEG with isotropic cubic Rashba SOC has been studied recently in detail by Karwacki \textit{et al.}~\cite{Karwacki2018Jun}. Here, we will use the general results for current-induced spin polarization presented in \cite{Karwacki2018Jun}, and adapt them to derive the theoretical description of BMR.

\subsection{Model assumptions}
Without losing the generality in our further analysis the external electric field is applied in the $x$-direction. This means that current-induced spin polarization, under zero magnetic field, has only $y$-component. 
We focus on the (unidirectional) bilinear correction to the magnetoresistance, i.e., we characterize the term in magnetoresistance that is proportional to the in-plane magnetic field and simultaneously to the charge current density (electric field). The higher order terms with respect to $B$ and $j$, that can eventually appear and lead to additional corrections in magnetoresistance (e.g., in strong magnetic fields) are not considered here. Accordingly, we treat perturbatively the terms proportional to the in-plane magnetic field and spin-orbital field in Hamiltonian~(\ref{eq:hamiltonian}), i.e., $H = H_0 + H_{pert}$, where:

\begin{equation}\label{eq:H0}
	H_0=\frac{\hbar^2k^2}{2m^{\ast}} \sigma_0 + i \alpha (k^3_-\sigma_+ - k^3_+\sigma_-)  ,
\end{equation}
and the perturbation is defined as:
\begin{equation}
\begin{aligned}
	H_{\mathrm{pert}} =\,& 
   \sum_{\beta = 0, x,y}   b_{\mathrm{eff} \beta}\, \sigma_\beta
\end{aligned}
\end{equation}
where 
\begin{equation}
b_{\mathrm{eff}\, 0} = B_x s^{x}_{0} + (B_y +J_{0} S_{y}) s^{y}_{0}
\end{equation}
\begin{equation}
b_{\mathrm{eff}\, x} = B_x s^{x}_{x} +  (B_y + J_{1} S_{y}) s^{y}_{x}
\end{equation}
\begin{equation}
b_{\mathrm{eff}\, y} = B_x s^{x}_{y} + (B_y + J_{1} S_{y})  s^{y}_{y} 
\end{equation}
Thus, the $H_{\mathrm{pert}}$ has a form of Zeeman-like term acting in the pseudospin space.

Note, that the effective spin-orbital field is approximated in our calculations by the non-equilibrium spin polarization under zero magnetic field. This simplification is justified as the additional components of spin polarization that appear in the presence of an external magnetic field are a few orders of magnitude smaller than the component which survives under zero magnetic field.  The magnetic field contribution to the leading term of spin polarization results in higher order terms in magnetoresistance that are neglected here. 

%------------------------------------
\subsection{Method}
%------------------------------------
To obtain magnetotransport characteristics, we have used Matsubara-Green's function formalism within linear response theory~\cite{mahan} and used the following formula for the quantum-mechanical average value of the observable, $O_{n}$ corresponding to the operator $\hat{O}_{n}$:
{\small{
\begin{eqnarray}
\label{eq:O_omega}
	O_{\alpha} = - \lim_{\omega \to 0} \frac{eE_\beta \hbar}{(2\pi)^3 \omega} \int d^2\textbf{k} \int d \varepsilon f(\varepsilon)
{ \mathrm{Tr}} \left\{\hat{O}_\alpha G_{\textbf{k}}^R(\varepsilon+\omega) \hat{v}_\beta \Delta_{\mathbf{k}}^{RA}(\varepsilon) 
 \right.  \nonumber \\
+ \left. 
 \hat{O}_\alpha \Delta_{\mathbf{k}}^{RA}(\varepsilon) \hat{v}_\beta G_{\textbf{k}}^A(\varepsilon-\omega)\right\} \hspace{0.5cm}
\end{eqnarray}
}}
where $\Delta_{\mathbf{k}}^{RA}(\varepsilon) = \left[G_{\textbf{k}}^R(\varepsilon)-G_{\textbf{k}}^A(\varepsilon)\right]$ and $G_{\textbf{k}}^{R/A} (\varepsilon)$ is  the retarded/advanced Green function related to the Hamiltonian $\hat{H}_{0}$. Derivation of this formula can be found, e.g., in ~\cite{mahan,Dyrdal2016Jul,Karwacki2018Jun}. The operator corresponding to $n$-th component of the charge current density is
\begin{equation}
\hat{j}_{\alpha} = e \hat{v}_{\alpha} = \frac{e}{\hbar} \frac{\partial \hat{H}_{0}}{\partial k_{\alpha}},
\end{equation}
and  spin-polarization should be derived using the spin operators given by Eq.~(\ref{eq:spinOp}).

%==========================================================
\section{Spin-to-charge interconversion: limit of zero magnetic field}
\label{sec:Spin-to-charge_B0}
%==========================================================

In the first step we consider 2DEG with cubic Rashba SOC under zero magnetic field. Based on Eq.~(\ref{eq:O_omega}) the general relation between dc charge current density and spin polarization can be derived. This relation is starting point for our further consideration of nonlinear magnetotransport.

%----------------------------------------
\subsection{Charge current density}
%----------------------------------------
In the dc limit, the charge current density, calculated based on Eq.~(\ref{eq:O_omega}), takes the following form:
\begin{equation}
\label{eq:jx_Tfinite}
\begin{aligned}
	j_x =& -\frac{e^2\hbar E_x}{4\pi\Gamma_0}
		\int d k \Bigg\{ \frac{3\alpha k^4}{m} [f'(E_+) - f'(E_-)]\\
		&+ \frac{k^3 \hbar^2 }{2m^2} [f'(E_+) + f'(E_-)]\\
		&+ \frac{9\alpha^2 k^5 }{2 \hbar^2} \left(1+ \frac{1 }{\left( \alpha k^3/\Gamma \right)^2 + 1} \right) [f'(E_+) + f'(E_-)] \Bigg\}
\end{aligned}
\end{equation}
In the low-temperature limit, the above expression leads to:
\begin{eqnarray}
\label{eq:jx_Tzero}
	j_x = \frac{e^2 \hbar E_x}{4 \pi \Gamma} \Bigg\{ \frac{9 \alpha^2 }{2\hbar^2} \left(k_{F-}^4 \nu_- + k_{F+}^4 \nu_+ \right) \hspace{2.4cm}\nonumber\\
		+ \frac{3 \alpha}{m} \left(k_{F+}^3 \nu_+ - k_{F-}^3 \nu_- \right) +\frac{\hbar^2
   }{2 m^2} \left(k_{F-}^2 \nu_-+k_{F+}^2 \nu_+\right)\nonumber \\
   	+ \frac{9\alpha^2}{2\hbar^2} \left(\frac{k_{F-}^4 \nu_-}{\left(\alpha k_{F-}^3/\Gamma\right)^2 + 1}+\frac{k_{F+}^4 \nu_+}{\left(\alpha k_{F+}^3/\Gamma\right)^2+1}\right) \Bigg\}\hspace{0.5cm}
\end{eqnarray}
where $k_{F\pm}$ are the Fermi wavevectors~\cite{Schliemann2005Feb}:
\begin{eqnarray}
\label{eq:FermiWavevectors}
k_{F\pm} = {\small{\mp \frac{1}{2}\frac{\hbar^2}{2m\alpha}
\left(
1 - \sqrt{1-4\pi n \left( \frac{2m\alpha}{\hbar^2} \right)}
\right)}}
\nonumber \hspace{2.5cm}\\
{\small{+ \left[
- \frac{1}{2} \left( \frac{\hbar^2}{2m\alpha}\right)^2
\left(
1 - \sqrt{1-4\pi n \left( \frac{2m\alpha}{\hbar^2} \right)}
\right)
+ 3\pi n 
\right]^{1/2}}}\hspace{0.5cm}
\end{eqnarray}
with $n$ denoting the charge carrier density and $\nu_\pm$ having sense of density of states:
\begin{equation}
\label{eq:nu}
	\nu_\pm = \frac{m}{\hbar^2 \left( 1 \pm \frac{3\alpha m}{\hbar^2}k_{F\pm} \right)}
\end{equation}
In the presence of randomly distributed point-like impurities, the relaxation rate, $\Gamma = \frac{\hbar}{2 \tau}$ ($\tau$ is the relaxation time), takes the form  $\Gamma =n_iV_0^2 \left(\nu_+ + \nu_- \right)/4$ where $n_i$ is the concentration of impurities and $V_{0}$ is the impurity potential. 

\subsection{Current-induced spin polarization}

The non-equilibrium spin polarization, $S_y$, in dc limit takes the form:
{\small{
\begin{equation}
\begin{aligned}
	S_y =& -\frac{e\hbar E_x}{4\pi}
		s_0 \int dk \Bigg\{ \frac{3\alpha k^4}{2\Gamma} [f'(E_+) - f'(E_-)]\\  
			&\hspace{2.8cm}+ \frac{k^3\hbar^2}{2m\Gamma} [f'(E_+) + f'(E_-)]\Bigg\}\\
		&- \frac{e\hbar E_x}{4\pi}
		s_1 \int dk \Bigg\{ \frac{k^4 \hbar^2}{2m\Gamma} [f'(E_+) - f'(E_-)]\\
   &+ \frac{3\alpha}{2} k^5 \left( \frac{1}{\Gamma} + \frac{1 }{(k^6 \alpha^2 + \Gamma^2)} \right) [f'(E_+) + f'(E_-)]
   \Bigg\},
\end{aligned}
\end{equation} 
}}
that in the low-temperature limit can be written as:
{\small{
\begin{eqnarray}\label{eq:Sy0_T0}
	S_y = \frac{e\hbar E_x}{8\pi\Gamma}
		s_0 \Bigg\{ \frac{\hbar^2}{m}\left(k_{F+}^2\nu_+ + k_{F-}^2\nu_-\right)  + 3\alpha \left(k_{F+}^3\nu_+ - k_{F-}^3\nu_-\right) \Bigg\} \nonumber \\
		+ \frac{e\hbar E_x}{8\pi\Gamma}
		s_1 \Bigg\{ 3\alpha \left(k_{F+}^4\nu_+ - k_{F-}^4\nu_-\right)  + \frac{\hbar^2}{m} \left(k_{F+}^3\nu_+ + k_{F-}^3\nu_-\right)\nonumber \\
		+ 3\alpha \left[ \frac{k_{F+}^4\nu_+}{\left( \alpha k_{F+}^3/\Gamma \right)^2+1} + \frac{k_{F-}^4\nu_-}{\left( \alpha k_{F-}^3/\Gamma \right)^2+1} \right] \Bigg\}.\hspace{0.8cm}
\end{eqnarray}
}}
%
%%%%%%

\noindent Note that in the above expressions $s_{0,1}$ are material parameters, that  characterize the  material and define the spin operators. Their explicit forms are provided in~\ref{App:AppA}.  Combining~(\ref{eq:jx_Tzero}) and~(\ref{eq:Sy0_T0}) it is possible to express the spin polarization by the charge current density:
\begin{equation}\label{eq:Sy0jx}
	S_y = \frac{s_0\mathcal{S}_{s0} - s_1\mathcal{S}_{s1}}{e\xi} j_x
\end{equation}
where $\xi = \frac{3\alpha}{\hbar^2} \left(k_{F+}^3-k_{F-}^3\right) + \frac{1}{m} \left( k_{F+}^2 + k_{F-}^2 \right)$. 
\\

In the limit of $\gamma_{1} \gg \gamma_2$ one finds:
\begin{equation}\label{eq:Sy_s0}
S_y = s_0 \frac{m}{e}j_x,
\end{equation}
where
\begin{equation}\label{eq:jx_s0}
j_{x} = \frac{e^{2}}{4\pi} \xi \tau E_{x}.
\end{equation}
with $\tau$ denoting relaxation time linked to the relaxation rate, $\Gamma$, through the simple relation $\Gamma=\hbar/(2\tau)$.

%=======================================
\section{Magnetoresistance}
\label{sec:magnetotransport}
%=======================================

%-----------------------------------
\subsection{Longitudinal charge conductivity in the presence of magnetic field}\label{sec:sigmaYY}
%-----------------------------------
The current-induced spin polarization determines the spin-orbital field. Treating the effective field $\mathbf{b}_{\mathrm{eff}}$ as a perturbation the diagonal conductivity can be expressed as: 
\begin{equation}
\label{eq:sigxx}
	\sigma_{xx} = \frac{j_{x}}{E_{x}} = \frac{e^2 \hbar}{2\pi} \int \frac{d^2\textbf{k}}{(2\pi)^2} \mathrm{Tr} \lbrace \hat{v}_x \bar{G}_{\mathbf{k}}^R \hat{v}_x \bar{G}_{\mathbf{k}}^A \rbrace
\end{equation}
where $\bar{G}_{\mathbf{k}}^{A/R}$ are impurity-averaged advanced and retarded Greens' functions in the weak magnetic field limit: $\bar{G}_{\mathbf{k}}^{R/A} = G_{\mathbf{k}}^{R/A} + G_{\mathbf{k}}^{R/A} H_{\mathrm{pert}} G_{\mathbf{k}}^{R/A} + G_{\mathbf{k}}^{R/A} H_{\mathrm{pert}} G_{\mathbf{k}}^{R/A} H_{\mathrm{pert}} G_{\mathbf{k}}^{R/A}$. 
Note that contributions related to two retarded and two advanced Green's functions (see Eq.~(\ref{eq:O_omega})) have been neglected, as they do not affect the final results.

\begin{figure}[t]
\includegraphics[width=.49\textwidth]{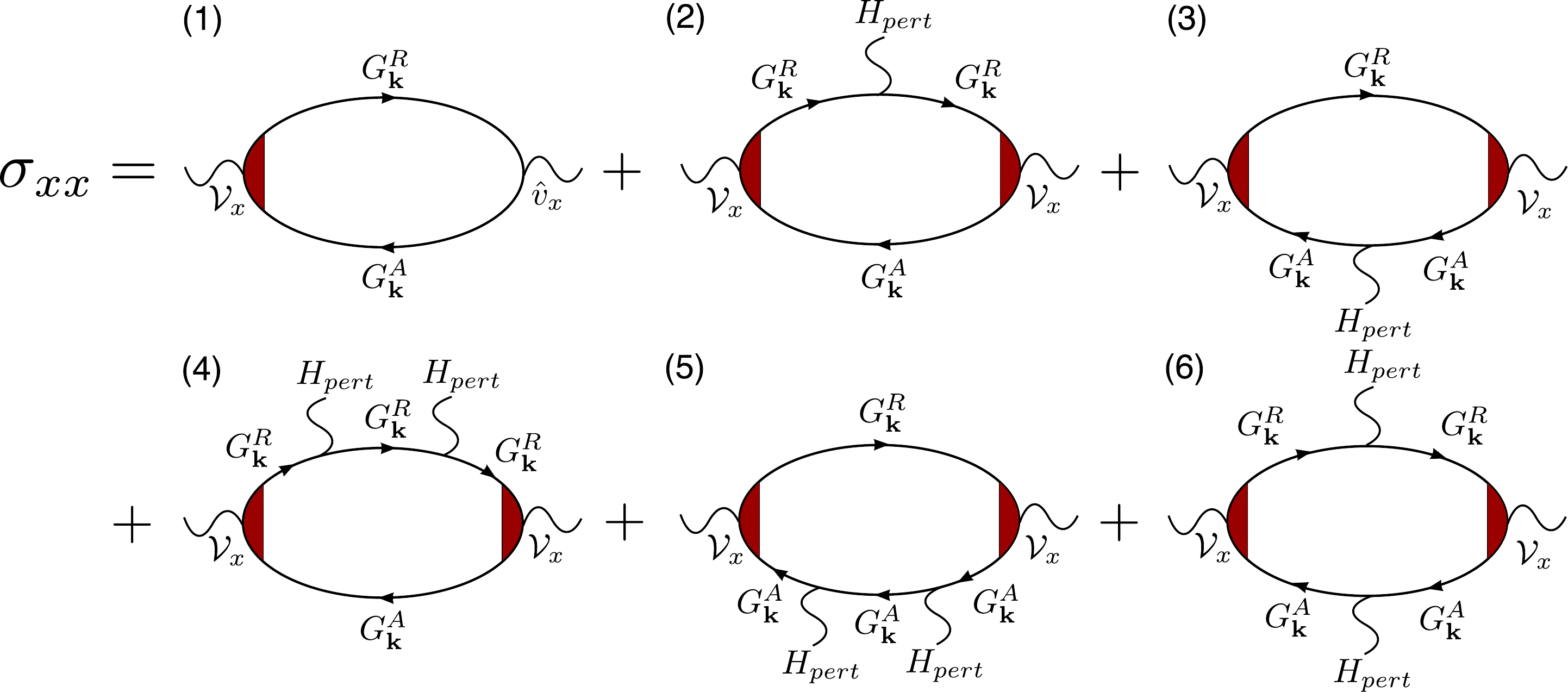}
	\caption{Feynman diagrams for diagonal conductivity up to the second order perturbation due to $\mathbf{b}_{\mathrm{eff}}$.}
\label{fig:FeymannDiagramsPert}
\end{figure}

Based on Eq.~(\ref{eq:sigxx}) and diagrammatic theory one finds six diagrams, depicted in Fig.~\ref{fig:FeymannDiagramsPert}, that can be grouped into three terms, according to the order of perturbation: 
\begin{equation}
    \sigma_{xx} = \frac{e^2 \hbar}{8\pi^3} \left( \mathcal{D}_{1} + \mathcal{D}_{2,3} + \mathcal{D}_{4-6}\right)
\end{equation}
The first diagram, related to the expression $\mathcal{D}_{1}$ describes the conductivity in the absence of $\mathbf{b}_{\mathrm{eff}}$, 
and is given by:

\begin{equation}\label{eq:sigmaxx1}
	\mathcal{D}_{1} = 
	\pi^2 \left[\frac{\xi}{\Gamma}+\frac{9 \alpha^2 \Gamma}{\hbar^2}  \left(\frac{k_-^4 \nu_-}{\Gamma^2+\alpha^2 k_-^6}+\frac{k_+^4
   \nu_+}{\Gamma^2+\alpha^2 k_+^6}\right)\right],
\end{equation}
which in the low-impurity concentration limit, $\Gamma \rightarrow 0$, leads to:
\begin{equation}\label{eq:sigmaxx1_Gamma0}
	\mathcal{D}_{1} = \pi^2\frac{\xi}{\Gamma}.
\end{equation}
%The correction to $\Gamma$ from effective field  is negligible in the limit $E_{3R} \gg E_{B}$. Thus, in the following  we approximate $\Gamma$ as $\Gamma_0$.
The second and the third diagrams do not contribute, as ${\mathcal{D}_{2,3} = 0}$.
The diagrams ${(4)-(6)}$ lead to  $\mathcal{D}_{4-6}$ that contains $B$-linear and $B$-quadratic terms:
\begin{equation}\label{eq:D46}
    \mathcal{D}_{4-6} = \left[B^2 \left(
	   \mathcal{F}_1^{B^2} + \mathcal{F}_2^{B^2}\cos(2\psi)\right)
	   + B \sin(\psi) S_y \mathcal{F}^{jB}
	   \right],
\end{equation}
with $\mathcal{F}$s being rather cumbersome functions of $k_\pm$, $\nu_\pm$ and $\alpha$, thus not shown here. 
%in the ~\ref{App:AppC} we provide explicit form for only $\mathcal{F}^{jB}$. 
%App (explicit forms are listed in the~\ref{App:AppC}).

%-----------------------------------
\subsection{Bilinear magnetoresistance}\label{sec:BMR}
%-----------------------------------
The magnetoresistance can be expressed in a standard form 
$\mathrm{BMR} = (\rho_{xx} - \rho_{xx}^{\scriptscriptstyle{0}})/\rho_{xx}^{\scriptscriptstyle{0}}$, where ${\rho_{xx}^{\scriptscriptstyle{0}}=1/\sigma_{xx}^{\scriptscriptstyle{(1)}}}$ is the diagonal resistance in the absence of magnetic field and ${\rho_{xx} = 1/(\sigma_{xx}^{\scriptscriptstyle{(1)}}+ \sigma_{xx}^{\scriptscriptstyle{(4-6)}})}$. Note that the transverse conductivity, $\sigma_{xy}$, (planar Hall effect) is much smaller than $\sigma_{xx}$ and has been neglected here. 
 
The unidirectional (bilinear) contribution to magnetoresistance is defined as: 
\begin{equation}
	\mathrm{BMR} = \frac{1}{2} \left[\mathrm{MR}(j_x=j) - \mathrm{MR}(j_x=-j)\right].
\end{equation}
Since the effect of in-plane magnetic field is assumed to be small, i.e., ${\mathcal{D}_{1} \gg \mathcal{D}_{4-6}}$,\, the diagonal resistance can be written\, as  ${\rho_{xx} = \rho_{xx}^{\scriptscriptstyle{0}} (1 - \mathcal{D}_{4-6}/\mathcal{D}_{1})}$, and BMR can be described as $\mathrm{BMR} = - \left[\mathcal{D}_{4-6}(j_x = +j) - \mathcal{D}_{4-6} (j_x = -j) \right]/(2 \mathcal{D}_1)$.
Finally, the BMR can be written in the form:
\begin{equation}\label{eq:BMR}
\begin{aligned}
	\mathrm{BMR} = -\frac{j_x B\sin(\psi)}{e \xi^2}
	(s_0\mathcal{S}_{s0} - s_1\mathcal{S}_{s1})\hspace{2.5cm}\\
	\times\left[J_0 s_0^{2} \mathcal{C}_{s0} + (J_0 +J_1) s_0 s_1 \mathcal{C}_{s0s1} 
% \right.\\
%    &\left. 
    + J_1 s_1^{2} \mathcal{C}_{s1})\right],
\end{aligned}
\end{equation}
where 
\begin{subequations}\label{eq:CParts}\small
\begin{align}
	 \mathcal{C}_{s0} 
	=\,&  \left[
	\frac{15}{2\hbar^2}(\nu_++\nu_-)
	-\frac{3\hbar^2}{2 m^2}\left(\nu_+^3 + \nu_-^3\right)
	-\frac{9\alpha}{4m}\left(k_+ \nu_+^3 - k_- \nu_-^3\right)
		\right]\\
	\mathcal{C}_{s0s1} 
	=\,& \left[-\frac{6}{\alpha \hbar^2}
	-\frac{9\alpha}{4 m}\left(k_-^2 \nu_-^3+k_+^2 \nu_+^3\right)
\right.\nonumber\\
	&\left.
\hspace{1.8cm} +\frac{9}{4m}\left(k_+ \nu_+^2-k_- \nu_-^2\right)
	+\frac{3}{\alpha m}(\nu_-+\nu_+)\right]\\
	\mathcal{C}_{s1} 
	=\,&  \left[\frac{9 \alpha}{4 m}\left(k_+^3 \nu_+^3-k_-^3 \nu_-^3\right)
	-\frac{3}{m}\left(k_-^2 \nu_-^2+k_+^2 \nu_+^2\right)\right.\nonumber\\
	&\left.
 \hspace{1.7cm}+\frac{2}{\alpha \hbar^2}(k_+-k_-)
	+\frac{29}{4\alpha m}(k_- \nu_--k_+ \nu_+)\right]
 \end{align}
 \end{subequations}
Eq.~(\ref{eq:BMR}) is the main result of this article.

%-------------------------------------
\subsection{The limit: $\gamma_1 \gg \gamma_2$}
%-------------------------------------

When $\gamma_1 \gg \gamma_2$, the coefficients $s_{0}\gg s_1$ and $J_{0}\gg J_1$. In such a case the spin polarization and charge current density are given by Eqs.~(\ref{eq:Sy_s0}), (\ref{eq:jx_s0}), whereas the spin-orbital field  $\mathbf{B}_{\mathrm{so}}\approx J_{0} S_y$. The expression describing bilinear magnetoresistance simplifies to 
\begin{equation}\label{eq:BMR_s0}
\begin{aligned}
	\textrm{BMR} = -\frac{j_x B\sin(\psi)}{e \xi^2}
	& J_0 s_0^3 \mathcal{S}_{s0} \mathcal{C}_{s0}.
\end{aligned}
\end{equation}
Taking explicit forms of $J_{0}$, $\mathcal{S}_{s0}$, $\mathcal{C}_{s0}$ we finally get the following formula:
\begin{equation}
\label{eq:BME_s0_F}
\begin{aligned}
\textrm{BMR} = \frac{3}{4} \pi s_0 \frac{\hbar}{e} \frac{\eta}{\xi^{2}} j_x B\sin(\psi)
\end{aligned}
\end{equation}
where
$\eta = \frac{\hbar^{2}}{m} \left(10 (\nu_{-} + \nu_{+}) + \frac{\hbar^{2}}{m} (\nu_{-}^{2} + \nu_{+}^{2}) - \frac{\hbar^{4}}{m^{2}} (\nu_{-}^{3} + \nu_{+}^{3})\right)$.

\begin{figure*}[!t]
	\includegraphics[width=1\textwidth]{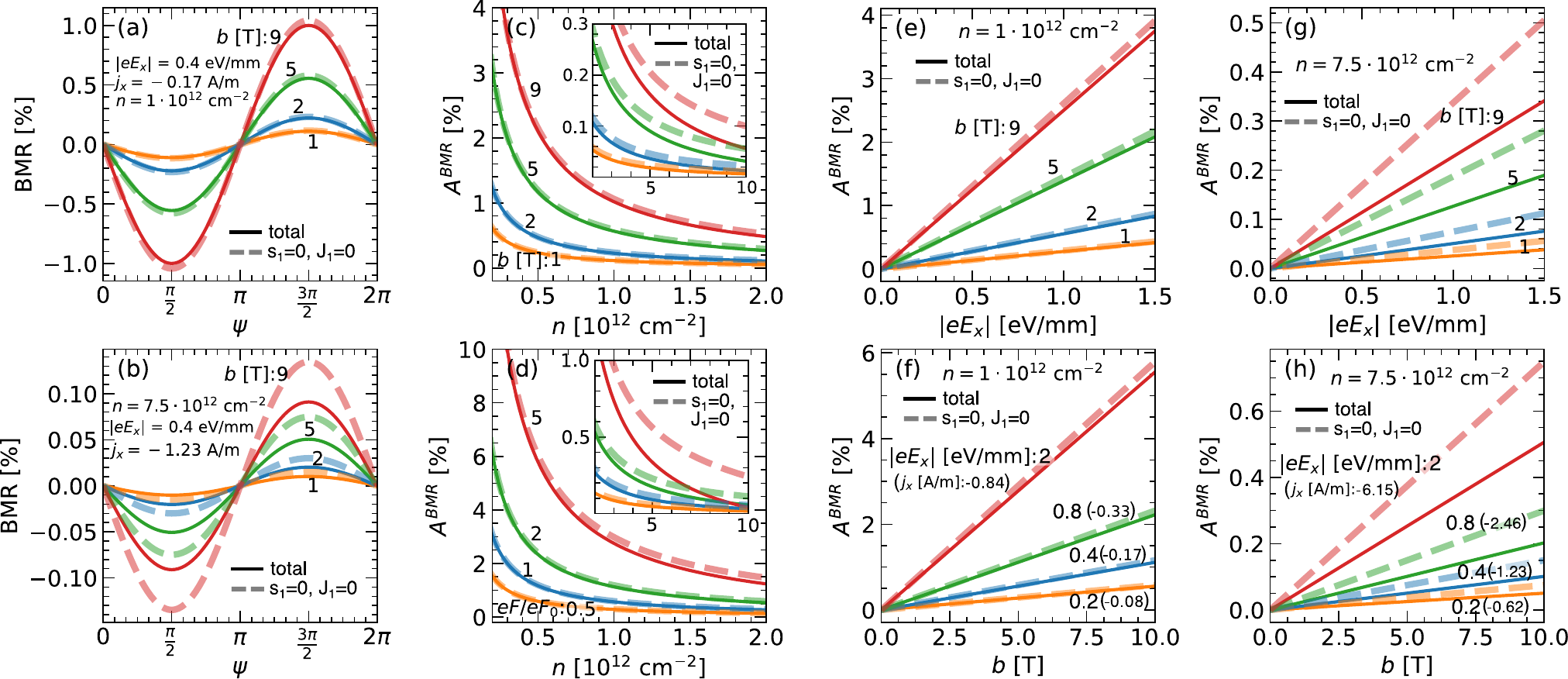}
	\caption{
 Bilinear magnetoresistance, $\mathrm{BMR}$, as a function of the angle $\psi$ (the angle between the in-plane magnetic field and $\hat{x}$-axis) for different values of magnetic field, $b$, (a,b).
 The amplitude of BMR, $A^{\mathrm{BMR}} = \mathrm{BMR}(\psi=3\pi/2)$, as a function of carrier density, $n$, for the indicated values of  magnetic field, $b$ (c) and for the indicated values of the quantum well  potential, $eF/eF_0$, (d).
 $A^{\mathrm{BMR}}$ as a function of the external electric field, $eE_x$, for the indicated values of magnetic field, $b$, and for  the carrier density $n= 1 \cdot 10^{12}$~cm$^{-2}$ (e) and $n= \{ 1, 7.5 \}\cdot 10^{12}$~cm$^{-2}$ (g). 
 $A^{\mathrm{BMR}}$ as a function of the in-plane magnetic field, $b$, for the indicated values of  electric field, $eE_x$, and for the  carrier density $n= 1 \cdot 10^{12}$~cm$^{-2}$ (f) and $n= \{ 1, 7.5 \}\cdot 10^{12}$~cm$^{-2}$ (h). 
 The solid lines represent the total formula on $\mathrm{BMR}$, Eq.~(\ref{eq:BMR}), while the light dashed lines correspond to the formula in the limit $\gamma_1\gg\gamma_2$, Eq.~(\ref{eq:BMR_s0}), i.e., for $s_1=0$ and $J_1=0$.
 Other parameters are taken from Refs.~\cite{Nakamura2012May,Karwacki2018Jun} for an STO-surface, i.e.,: $eF_0=450\cdot10^6$~eV/m, $\gamma_1=0.66$, $\gamma_2=0.003$, $L_z=100\cdot 10^{-10}$~m, $m=1.5m_0$ and $J_1/J_0 \cong -0.23$.}\label{fig:BMRy}
\end{figure*}
%----------------------------------
\subsection{Numerical results}
%----------------------------------

%\textcolor{ao}{\textit{frankly, all the formulas are analytical}}

Figure~\ref{fig:BMRy} collects numerical results, obtained based on formula~(\ref{eq:BMR}).
Figures~\ref{fig:BMRy} (a),(b) present the BMR as a function of the angle $\psi$ defining the relative orientation of charge current density and external in-plane magnetic field  (i.e., $\psi$ defines the orientation of in-plane magnetic field with respect to the $\hat{x}$-axis). Linear dependencies of the BMR signal amplitude, $A^{\mathrm{BMR}}$, as a~function of the amplitude of magnetic field, $b$, and electric field, $eE_x$, are presented in Figs.~\ref{fig:BMRy}(e)-(h). Finally Fig.~\ref{fig:BMRy}(c) shows $A^{\mathrm{BMR}}$ as a function of the carrier concentration, $n$, for the fixed amplitude of the in-plane magnetic field $b$, whereas Fig.~\ref{fig:BMRy}(d)  presents $A^{\mathrm{BMR}}$ as a function of $n$ for the fixed strength of quantum well potential,  $eF$. The presented results reflect the linear dependence of BMR with respect to both electric and magnetic fields. Moreover, for the low carrier density, the simplified expressions (29), (30) and (39) are sufficient to describe the charge current, spin polarization, and BMR. In Fig.~\ref{fig:BMRy}, the results obtained from the simplified expressions are indicated by the dashed lines, whereas the solid lines present the BMR described by the full expression (37).  For higher charge carrier concentrations, one can note a deviation of the results obtained form the simplified expression from those based on the full formula. In turn, the BMR signal decreases with increasing $n$. Accordingly,  the simplified expression is sufficient in the carrier concentration range for which one can expect the strongest BMR signal. 

%------------------------------
\section{Summary}
\label{sec:summary}
%-------------------------------
We have studied theoretically the unidirectional magnetoresistance in a 2D electron gas with the isotropic cubic form of Rashba spin-orbit coupling under external in-plane magnetic field. The mechanism leading to the unidirectional magnetoresistance response is here based  on the effective spin-orbital field originated from non-equilibrium spin polarization.
The obtained analytical results indicate that BMR can be of an order of a few percent in a range of low carrier densities. The results presented in the manuscript have been derived for the model Hamiltonian that may be used for the description of electronic properties of p-doped semiconductor quantum wells as well as for the description of electron gas emerging at  surfaces or interfaces of perovskite oxides (for a certain energy window). Thus, these results may be useful for interpretation of experimental data of a relatively large group of materials. For example, the bilinear magnetoresistance has been measured recently in LAO/STO~\cite{Dyrdal2020Jan}, where a new scheme for determining the linear Rashba parameter has been proposed. The theoretical model provided in our article allows determining the cubic form of Rashba coupling parameter in such structures, with chemical potential gated above the Lifshitz point, where energy subbands reveal strong anisotropy due to cubic Rashba term.

%----------------------------------
\section*{Acknowledgement}
%----------------------------------

\noindent This work has been supported by the National Science Centre in Poland - project NCN Sonata-14, no.: 2018/31/D/ST3/02351.

%----------------------------------
\appendix
%----------------------------------

%---------------------------
\section{Linking the effective  Hamiltonian~(\ref{eq:hamiltonian}) to the materials parameters}
\label{App:AppA}
This appendix collects explicit forms of the parameters and spin operators corresponding to Hamiltonian~(\ref{eq:hamiltonian}). Accordingly, the effective mass takes the form:
\begin{eqnarray}
m^{\ast} = m_{0} \left( \gamma_1 + \gamma_2 - \frac{256 \gamma_{2}^{2}}{3\pi^2(3 \gamma_{1} + 10 \gamma_2)}\right)^{-1}, 
\end{eqnarray}
with the electron rest mass $m_{0}$, and $\gamma_{1,2}$ denoting phenomenological parameters of the Luttinger Hamiltonian~\cite{LuttingerKohn1955,Luttinger1956}.

The Rashba coupling parameter, $\alpha$ is defined by the following expression:
\begin{equation}\label{eq:alpha}
\alpha = \frac{512 e F L_{z}^{4} \gamma_{2}^{2}}{9 \pi^6 (3 \gamma_{1} + 10 \gamma_{2}) (\gamma_{1} - 2 \gamma_{2})}
\end{equation}
where $L_{z}$ and $eF$ are the width and potential of the quantum well, respectively.

Hamiltonian (1) has been obtained perturbatively based on $8\times8$ Luttinger Hamiltonian and mapping into the lowest heavy-hole subbands~\cite{Liu2008Mar}, thus, spin operators are no longer simply expressed by Pauli matrices $\sigma_{n}$. To obtain proper spin operators one needs to perform the same unitary transformations as that made to obtain Hamiltonian (1)~\cite{Liu2008Mar}. The explicit forms of spin operators are listed below: 
\begin{equation}
\hat{s}_{\alpha} = \sum_{\beta = 0,x,y,z}s^{\alpha}_{ \beta} \sigma_\beta ,
\end{equation}
${s^{x}_{0} = -s_0 k_y}$,\,  ${s^{y}_{0} = s_0 k_x}$,\, ${s^{x}_{x} = s^{y}_{y} = s_1 (k_x^2 - k_y^2)}$, ${s^{x}_{y} = 2 s_1 k_x k_y}$,  ${s^{y}_{x} = - s^{x}_{y}}$, $s^{z}_{0}=s^{z}_{x}=s^{z}_{y}=0$, $s^{z}_{z} = 3 \hbar/2$.
The parameters $s_{0,1}$ are material parameters defined as follows:
\begin{equation}
 s_0 = \frac{512eFL_z^4\gamma_2 m_0}{9\pi^6 \hbar^2(3\gamma_1+10\gamma_2)(\gamma_1-2\gamma_2)}
\end{equation}
\begin{equation}
 s_1 = \left( \frac{3}{4\pi^2} - \frac{256\gamma_2^2}{3\pi^4 (3\gamma_1+10\gamma_2)^2}\right) L_z^2 .
\end{equation}

It should be stressed that in the energy window for which the considered model Hamiltonian is applicable (i.e., small carriers concentration) one finds  $\gamma_1 \gg \gamma_2$ and contribution to the spin-dependent transport properties proportional to $s_1$ is rather small, and can be negligible (see for example data collected in Tab.~I in~\cite{Karwacki2018Jun}).
In such a case:
\begin{equation}
\label{eq:spinOperators_s0}
\hat{s}_{x} \approx - s_0 k_y \sigma_0,  \quad \hat{s}_{y} \approx s_0 k_x \sigma_0, \quad \hat{s}_{z} =  \frac{3}{2} \sigma_z 
\end{equation}
\begin{equation}
H_{\mathrm{pert}} = b_{\mathrm{eff}\, 0}\, \sigma_0
\end{equation}
In this paper, we consider magnetoresistance for the general effective Hamiltonian, as well as for the simplified model that leads to simpler analytical expressions applicable in the limit $\gamma_1 \gg \gamma_2$.

%------------------------------------
\section{Estimation of $J_{0,1}$ parameter}
\label{App:Jparameters}
%------------------------------------

Under an external electric field, the Fermi contour is shifted in the momentum space by $\Delta k_{x} = - \frac{e \tau}{\hbar} E_x$. In turn, the diagonal charge current density is given by eq.~(\ref{eq:jx_s0}) (the leading term), thus one finds the relation between $\Delta k_x$ and $j_x$ in the form ${\Delta k_x = - \frac{4\pi}{e \hbar \xi} j_{x}}$.

Due to the correction originating from the non-equilibrium shift in the momentum space, the Hamiltonian $H_0$, Eq.~(\ref{eq:H0}), takes the form $H_{0} (k_x + \Delta k_{x})  = H_{0} + H_{\Delta k_x}$, where:
\begin{eqnarray}
\label{eq:HDk}
 H_{\Delta k_x}  =
\frac{\hbar^{2}}{2 m} k_x \Delta k_{x} \sigma_0 + 6 \alpha k_x k_y \Delta k_{x} \sigma_{x}\hspace{2cm} \nonumber\\ - 3 \alpha(k_x^2 - k_y^2) \Delta k_{x} \sigma_{y} + \it{O}({\small{(\Delta k_{x})}}^2) .\hspace{0.5cm}
\end{eqnarray}
In turn, the effective spin-orbital field  introduced to our theory in (\ref{eq:hamiltonian}) is defined as:
\begin{eqnarray}
\label{eq:JSy}
B_{so\,y}\hat{s}_{y} = J S_{y} \hat{s}_{y} = J S_{y} s_0 k_x \sigma_{0} - J S_y 2 s_1 k_x k_y \sigma_x \nonumber\\
+ J S_{y} s_1 (k_x^{2} - k_y^2) \sigma_y
\end{eqnarray}
By the comparison of terms standing in front of the same Pauli matrices in Eq.~(\ref{eq:HDk}) and Eq.~(\ref{eq:JSy}) one finds:
\begin{align}
J S_y s_0 &= \frac{\hbar^{2}}{2 m} \Delta k_x\\
- s_1 J S_y &= 3 \alpha \Delta k_x\\
s_1 J S_y  &= - 3 \alpha \Delta k_x
\end{align}
Accordingly, the above relations indicate that two coupling constants should be introduced. From the first equality, one finds $J = J_0$, whereas 
the second and third equalities are identical and result in $J = J_1$. Thus the parameters $J_{0,1}$ reads
\begin{equation}
\label{eq:J0}
J_{0} = - \frac{2\pi\hbar}{m s_0} (s_0 \mathcal{S}_{s0} - s_1 \mathcal{S}_{s1})^{-1} 
\end{equation}
\begin{equation}
\label{eq:J1}
J_1 = \frac{4\pi}{\hbar} \frac{3 \alpha}{s_1}  (s_0 \mathcal{S}_{s0} - s_1 \mathcal{S}_{s1})^{-1}
\end{equation}
where
\begin{subequations}\label{eq:xi_Ss01}
\begin{align}
    \mathcal{S}_{s0} =\,& 3\alpha \left(k_{F-}^3 \nu_- + k_{F+}^3 \nu_+ \right)
+ \frac{\hbar^2}{m} \left(k_{F-}^2 \nu_- + k_{F+}^2 \nu_+ \right)\nonumber\\
=\,& k_{F+}^2 + k_{F-}^2\\
    \mathcal{S}_{s1} =\,& 3\alpha \left(k_{F-}^4 \nu_- + k_{F+}^4 \nu_+ \right)
+ \frac{\hbar^2}{m} \left(k_{F-}^3 \nu_- + k_{F+}^3 \nu_+ \right)\nonumber\\
=\,& k_{F+}^3 - k_{F-}^3
\end{align}
\end{subequations}

\bibliography{cas-refs.bib}





% \end{thebibliography}
\end{document}